\documentclass[12pt,a4paper]{article}
\usepackage{epsfig}
\usepackage{cite}
\usepackage{mcite}
\usepackage{array,tabularx,epsfig,mathrsfs,graphicx,rotating}
\usepackage{ifthen}
\usepackage{amsfonts}
\usepackage{ragged2e}

\PassOptionsToPackage{hyphens}{url}
\usepackage[hyphens]{url}
\usepackage{hyperref}

\hypersetup{
  colorlinks=true,
  linkcolor=blue,
  citecolor=blue,
  urlcolor=blue
}

\usepackage{breakurl}

\newcommand{\beq}{\begin{equation}}
\newcommand{\eeq}{\end{equation}}

\chardef\til=126

\begin{document}

\clearpage
\pagestyle{empty}
\setcounter{footnote}{0}\setcounter{page}{0}%
\thispagestyle{empty}\pagestyle{plain}\pagenumbering{arabic}%

\hfill  ANL-HEP-CP-13-32
 
\hfill June 20,  2013

\hfill Version 1.0 

\vspace{2.0cm}

\begin{center}

%%%%%%%%%%%%%%%%%%%%%%%%%%%%%%%%%%%%%%%%%%%%%%%%%%%%%%%%%%%%%%%
{\Large\bf
Next generation input-output data format \\ for HEP using Google's protocol buffers
\\[-1cm] }
%%%%%%%%%%%%%%%%%%%%%%%%%%%%%%%%%%%%%%%%%%%%%%%%%%%%%%%%%%%%%%%

\vspace{2.5cm}

{\large S.V.~Chekanov 

\vspace{0.5cm}
\itemsep=-1mm
\normalsize
\small
HEP Division, Argonne National Laboratory,
9700 S.Cass Avenue, \\ 
Argonne, IL 60439
USA
}

\normalsize
\vspace{1.0cm}

% put line numbers
% \linenumbers

\vspace{0.5cm}
\begin{abstract}
We propose a data format for Monte Carlo (MC) events, or any structural data, 
including experimental data, in a compact binary form using variable-size integer encoding 
as implemented in the Google's Protocol Buffers package.
This approach is implemented in the so-called {\sc ProMC} library which  
produces smaller file sizes for MC records compared to the
existing input-output libraries used in high-energy physics (HEP).
Other important features are
a separation of abstract data layouts from concrete programming implementations, self-description
and random access. Data stored in {\sc ProMC} files 
can be written, read and manipulated 
in a number of programming languages, such  C++, Java and Python.
\end{abstract}

\end{center}

\newpage
\setcounter{page}{1}

% put line numbers
% \linenumbers

%%%%%%%%%%%%%%%%%%%%%%%%%%%%%%%%%%%%%%%%%%%%%%%%%%%%%%%%%%%%%%%%%%
\section{Introduction}
%%%%%%%%%%%%%%%%%%%%%%%%%%%%%%%%%%%%%%%%%%%%%%%%%%%%%%%%%%%%%%%%%%

A crucial requirement for many scientific applications is store, 
retrieve and process large-scale numeric data with a small signal and large 
background (or ``noise''). Information on background objects is not required to be 
stored with the same numeric precision as that for signal objects. 
An input/output library which dynamically compresses data depending on the content of information 
becomes important for effective data storage and analysis. 

	A typical example is  the Large Hadron Collider (LHC) experiments designed to 
investigate  proton-proton and heavy-ion collisions in order to understand the basic structure of matter. 
The LHC experiments are currently involved in event processing and physics analysis of petabytes of data.
A single analysis requires a processing of tens of terabytes of data 
located on the grid storage across the glob. 
For example, the  number of collisions recorded by the ATLAS experiment since 2009 exceeds 20 billion. 
The number of particles in a single collision  will increase by a factor 5-10 for future high-luminosity LHC runs. 
Currently, the LHC experiments store more than 100 petabytes of data and this number 
will increase by a factor 10 over the next 10 years. 
A large fraction of stored data has  
a small fraction of "signal" particles, while most of low-energetic particles from other events
are less interesting and represent ``pileup'' background.
It is important to store pileup particles to derive corrections, 
but to store such particles with the same numeric precision as signal particles is inefficient.

In this paper we will discuss an input-output library which has a content dependent compression. 
It stores less energetic particles with a reduced numerical precision and smaller numbers of bytes. 
The library is designed for Monte Carlo (MC) simulation events,
but it can naturally be extended to store any information. The library was created during the Snowmass Community Studies \cite{snowmass}
with the goal to store MC simulations in a compact form on public web pages.

\section{The proposal}
This paper discusses an input/output persistent  framework which:

\begin{itemize}
\item
streams data into a binary form and dynamically writes less interesting, 
low-energetic particles with reduced precision compared to more energetic “signal” particles. 
For example, we expect that such content-dependent compression will decrease the LHC data volume by 30\% or more.
Although we use the word ``compression'', it should be noted that no
standard compression algorithms (gzip, zip, bunzip2) are used since 
the file-size reduction is achieved using a highly efficient binary format.

\item
is multiplatform. Data records can be manipulated in C++, Java and Python. This opens the possibility 
to use a number of ``opportunistic'' platforms for data analysis, such as Windows or Android,  which have not been used widely 
in HEP.

\item
is a self-describing data format based on a template approach to encode complex data structures. 
One can generate C++, Java and Python analysis codes from a {\sc ProMC} file itself. 

\item
has random access capabilities. Events can be read starting at any index. Individual events can be accessed 
via the network without downloading the entire files. 
Metadata can be encoded for each record, allowing for a fast access to interesting events.

\item
is implemented as simple, self-containing library which can easily be deployed on a 
number of architectures including
supercomputers, such as IBM Blue Gene/Q system. 
\end{itemize}

The proposed input-output framework is expected to be used in many scientific areas. In particular, it is useful for
(a) data reduction for large general-purpose detectors at colliders and other experiments; 
(b) effective data preservation; 
(c) effective data analysis without CPU overhead due to the standard decompression algorithms.

\section{Existing approaches}

The LHC experiments store data and experiment-specific MC events in compressed ROOT format \cite{root}. 
To store events generated by MC models in a more generic and exchangeable way, HEPMC \cite{hepmc}, STDHEP \cite{stdhep},
HepML \cite{Belov:2010xm} and the Les Houches event format (LHEF) \cite{Alwall:2006yp} 
file formats have been developed. 
For example, the  HEPMC library is interfaced with all major Monte Carlo models and is widely used by the HEP community
due to its simplicity,
platform independence, exchangeability and reusability.
However, the HEPMC format stores data in  uncompressed ASCII files, 
which are typically ten times larger than  ROOT files with the default compression. 

        The ROOT IO is an integrated part of the C++ ROOT analysis framework \cite{root} developed at CERN.
This framework is heavily integrated to the Linux platform.
Despite its popularity to store experiment-specific data,
it uses the  GZIP compression which is CPU intensive and lacks 
flexibility in store particles depending on their importance.
As discussed before,  the current and future LHC experiments will collect events with only a small fraction of
signal particles that are important for analyzers,  while most of low-energetic particles
from other (overplayed) events represent ``pileup'' background.
For high-luminosity LHC runs, one ``signal'' event (for example, event with a Higgs particle) will
contain 50-140 pileup events, with up to 10,000 low-energy particles that are not 
important for analysis of the signal itself. Still,  such particles (or a fraction of such particles) 
should be kept to derive corrections to signals.
Therefore, to store low-energetic particles with a smaller numeric precision becomes crucial in effective data
storage and analysis. The fixed-number of bytes to represent numeric data used by ROOT and by other data formats
does not allow to implement a compression that depends on particle content
(particle energy, mass, origin, etc.).
Another crucial aspects that can be improved is the ability to identify interesting events and
particles without reading entire files. This requires random access, effective metadata model, and the ability
to process files on multiple cores of high-performance multicore computers.

\section{Varint data encoding}

A possible solution for data reduction is to use
``varints'' which can encode integer (int32, int64 etc.) values using variable number of bytes.
Such algorithm is implemented in the Google's Protocol Buffers library \cite{protobuf}.
This library encodes complex data in the form of platform-neutral ``messages'' which 
can include the varint representation of data.
Smaller integer numbers represented by varints in such messages take a smaller number of bytes. 
For HEP applications, this implies that four-momenta of low-energetic particles  encoded using the integer representation
can be represented with a smaller number of bytes.
In addition, many particle
characteristics (such as particle status, particle ID, etc.) should be  represented by integer values 
anyway and thus are well suited to the varint representation.

	Although the varint data encoding  is available in the Protocol Buffers library 
publicly released by Google, this library alone is not sufficient to pursuit the goal of 
creating large files with multiple logically-separated records. 
The Protocol Buffers approach is most effective if each separate Protocol Buffers message has a
size less than 1MB (as recommended by Google). 
The  major problems that need to be addressed are: 1) to design Protocol Buffers message to store
particles in a single event to allow for the varint representation; 2) to find a method of serialization
of multiple messages (``events'') into a file which can keep many events; 3) how to implement metadata
model for fast access of interesting events and particles.
While (1) is rather specific to HEP, (2) and (3) are very general issues
that have to be solved in any research area where logically-separated event
records with varint-based  information is an attractive option for data
storage and processing. Because of such problems,
the usage of Google's Protocol Buffers to keep large numeric data is still limited in science and technology.

\subsection{Current implementation}
\label{cimp}

\begin{figure}
\begin{center}
\includegraphics[width=0.28\textwidth]{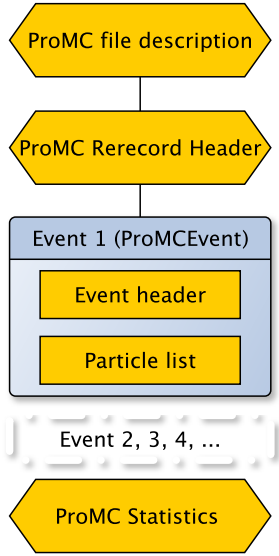}
\end{center}
\caption{
A schematic representation of the {\sc ProMC} file format. 
All records are encoded using the Protocol Buffers messages.
In addition, some metadata information is stored as text files inside the ZIP file for easy access on platforms without
installed {\sc ProMC} library.
}
\label{fig:graph}
\end{figure}

The {\sc ProMC} library \cite{promc} is designed to store HEP collision events
using the Google's Protocol Buffers library as a backend. 
The data are  stored in a file with the file headers and
multiple logically-separated messages.
Each separate message leverages the varint encoding for representing a single MC event.    
Figure~\ref{fig:graph} shows a schematic representation of a {\sc ProMC} file.
All Protocol Buffers  messages are stored as ZIP entries inside the {\sc ProMC} file.

To work with the {\sc ProMC} files, the {\sc ProMC} library does not need to be installed. 
This C++ library has to be installed
if events will be written or read in C++. {\sc ProMC} files can also be read and created using Java or Python,
without the platform-dependent {\sc ProMC} library.

The current {\sc ProMC} library is built on the assumption that the new  data format should be self-describing.
The current implementation allows to generate  analysis source codes (in C++, Java, Python) from a {\sc ProMC} file itself without knowing how
it was originally created. 
Another notable features are random access to any given event and a possibility to stream 
individual events through the network without reading or downloading entire files. 

Several benchmarks have shown that the {\sc ProMC} files are rather compact, typically 40\% smaller than ROOT files (See Table~1).
The {\sc ProMC} file size  depends on many factors, but there are two major factors that should be mentioned:
 1)  energy spectra of stored particles; 2) what exactly information is stored; 3) how the information 
is represented using integer values and what 
are their values. 
In the first case, particle spectra with large fraction of low-energy (``soft'') particles will be stored more effectively than particles 
with large values of four-momenta. 
The file size depends on a conversion factor which converts float values to integer representation.
The default mapping between the energy units used in HEP and int64 varints is given in Table~2. 
This layout can be changed since the conversion factor is included into the metadata section of the {\sc ProMC} file format.
In case of events with large pileup (i.e. a large fraction of soft particles), {\sc ProMC} files 
can lead to  almost a factor two smaller files compared to ROOT files with the same information  
based on the Double32\_t data types. 

The second factor that determines data reduction depends on the fraction of information which can be presented 
using integer values (int32, int64, etc.).
For a typical MC truth event record, several particle characteristics can be represented using 
varints without loosing numeric precision. The examples include particle ID,
status code, 1st and 2th daughter and mother particles. In many cases, their values are small 
and thus can be represented by only a few bytes.

\begin{table}
\begin{center}
    \begin{tabular}{lc}
    \hline
    File format  & File size (in MB)  \\ \hline 
    ASCII HEPMC file               & 346 \\
    ASCII HEPMC (after gzip compression)            & 138 \\
    ROOT file (using Double32\_t)       & 158 \\
    ProMC file                    & 112 \\
     \hline
    \end{tabular}
\caption{Typical file sizes for a benchmark included into the {\sc ProMC} package.}
\end{center}
\label{xtab1}
\end{table}

\begin{table}
\begin{center}
    \begin{tabular}{llc}
    \hline
    Energy  &  int64 representation & Nr of bytes  \\ \hline
    0.01 MeV      & 1  & 1   \\
    0.1  MeV      & 10  & 1   \\
    1    MeV      & 100  & 2   \\
    1    GeV      & 100 000  & 4   \\
    1    TeV      & 100 000 000  & 8  \\ 
    20   TeV      & 2000 000 000 & 8  \\ 
   \hline
    \end{tabular}
\caption{The default {\sc ProMC} 
mapping between energy units and int64 (varint) representation, together with the number of bytes used for the encoding.} 
\end{center}
\label{xtab2}
\end{table}

The {\sc ProMC} library includes several packages and tools to work with stored information:

\begin{itemize}
\item a Java browser to open {\sc ProMC} files in order to study  data layout of data, as well as the stored data.
Currently, the latter  feature supports only truth MC records. Files can be streamed using the network. 

\item
A number of tools to work with {\sc ProMC} files implemented in Python. Some of them are described below:

\begin{itemize}
\item {\it  promc\_info file.promc} - displays the information on the {\sc ProMC} file
\item {\it promc\_proto file.promc} - extracts the file layouts in the form of Protocol Buffers data templates
\item {\it promc\_code } - generates analysis code in C++, Java and Python. 
\item {\it promc\_log file.promc} - extracts a log file (if attached).
\item {\it promc\_extract file.prom  out.promc N} - extracts first ``N'' events and save them in the file ``out.promc'' 
\item {\it hepmc2promc file.hepmc file.promc ``description'' } - converts a HEPMC file \cite{hepmc} to the {\sc ProMC} file 
\item {\it stdhep2promc file.stdhep file.promc ``description'' } - converts a STDHEP file \cite{stdhep} to the {\sc ProMC} file
\end{itemize}

\end{itemize}

\subsection{Tutorials and examples}

The {\sc ProMC} library was used for the Snowmass 2013 community studies in order to 
store truth and {\sc DELPHES} \cite{Ovyn:2009tx} MC events in a compact binary form on the 
web servers for HEP community
The {\sc DELPHES} fast simulation can read the {\sc ProMC} files \cite{Ovyn:2009txd} with the truth records and convert
such files to reconstructed events. 
A number of examples illustrating how to read and write the {\sc ProMC} files is given on the {\sc ProMC} web page \cite{promc}.
The web page includes several examples of how to read, write and manipulate with {\sc ProMC} files 
using several programming languages: C++, Java and Python. Some basic information on stored data can also be extracted using PHP.

For C++/Python examples of reading and writing data, ROOT/PyROOT can be used for graphical visualization.
There are also examples which illustrate 
how to read data using Jython, the Python language implemented in Java. In this case, no platform specific libraries are used to read and display the data.
In  case of Java or Jython, {\sc SCaVis} \cite{scavis,Chekanov:1261772} and {\sc Jas} \cite{java_toni} Java-based analysis environments 
can be used for visual representation of data (histograms, scatter plots etc.). 
There is also an example which shows the random access capability of the {\sc ProMC} format 
and how to access certain events from a network without reading the entire {\sc ProMC} file.
In addition, a few examples are given which illustrate how to fill {\sc ProMC}
 files directly from the {\sc Pythia8} \cite{Sjostrand:2007gs} MC program or from HEPMC
files.

\section*{Acknowledgements}
I would like to thank P.~Van Gemmeren, J.Proudfoot and K.Strand for discussion. 
The submitted manuscript has been created by UChicago Argonne, LLC,
Operator of Argonne National Laboratory (``Argonne'').
Argonne, a U.S. Department of Energy Office of Science laboratory,
is operated under Contract No. DE-AC02-06CH11357.

%%%%%%%%%%%%%%%%%%%%%% references %%%%%%%%%%%%%%%%%%%%%%%%%%%%%%
\bibliographystyle{l4z_pl}
\def\bibname{\Large\bf References}
\def\refname{\Large\bf References}
\pagestyle{plain}
\bibliography{biblio}

\end{document}